# GRAPHENE: STATUS AND PROSPECTS

A. K. Geim

Manchester Centre for Mesoscience and Nanotechnology, University of Manchester,
Oxford Road M13 9PL, Manchester, UK

*Graphene is a wonder material with many superlatives to its name. It is the thinnest material in the universe and the strongest ever measured. Its charge carriers exhibit giant intrinsic mobility, have the smallest effective mass (it is zero) and can travel micrometer-long distances without scattering at room temperature. Graphene can sustain current densities 6 orders higher than copper, shows record thermal conductivity and stiffness, is impermeable to gases and reconciles such conflicting qualities as brittleness and ductility. Electron transport in graphene is described by a Dirac-like equation, which allows the investigation of relativistic quantum phenomena in a bench-top experiment. What are other surprises that graphene keeps in store for us? This review analyses recent trends in graphene research and applications, and attempts to identify future directions in which the field is likely to develop.*

Graphene research has developed a truly relentless pace. Several papers appear every day and, if the bibliometrics predictions (*1*) are to be trusted, the amount of literature on graphene will keep rapidly increasing over the next few years. This makes it a real struggle to keep up with the developments. Newcomers are left without a broad perspective and are largely unaware of previous arguments and solved problems, whereas the community's doyens already show signs of forgetting their earlier papers. To combat this curse of success, many reviews have appeared in the last two years, and books on graphene are in the making. Among those, let me recommend our review (*2*) that provides a good introductory material and is far from being obsolete. The electronic properties of graphene were recently discussed in an extensive theory review (*3*), and this basic information is unlikely to require any revision soon. Also, I recommend more specialized papers discussing the quantum Hall effect in graphene, its Raman properties, epitaxial growth on SiC, etc. which are collected in (*4*). Despite or, perhaps, because of the vast amount of available literature, graphene research has now reached the stage where a strategic update is needed to cover the latest progress, emerging trends and opening opportunities. This paper is intended to serve this purpose without repeating, whenever possible, the information available in the earlier reviews.

**Growing Opportunities**
Graphene is a single atomic plane of graphite, which – and this is essential – is sufficiently isolated from its environment to be considered free-standing. Atomic planes are of course familiar to everyone as constituents of bulk crystals but one-atom-thick materials such as graphene remained unknown. The basic reason for this is that nature strictly forbids the growth of low-dimensional (D) crystals (*2*). Crystal growth implies high temperatures (*T*) and, therefore, thermal fluctuations that are detrimental for the stability of macroscopic 1D and 2D objects. One can grow flat molecules and nm-sized crystallites, but as their lateral size increases, the phonon density integrated over the 3D space available for thermal vibrations rapidly grows, diverging on a macroscopic scale. This forces 2D crystallites to morph into a variety of stable 3D structures. The impossibility to grow 2D crystals does not actually mean that they cannot be made artificially. With hindsight, this seems trivial. Indeed, one can grow a monolayer inside or on top of another crystal (as an inherent part of a 3D system) and then remove the bulk at sufficiently low *T* such that thermal fluctuations are unable to break atomic bonds even in macroscopic 2D crystals and mold them into 3D shapes.

This consideration allows two principal routes for making 2D crystals (Fig. 1). One is to mechanically split strongly-layered materials such as graphite into individual atomic planes (Fig. 1A). This is how graphene was first isolated and studied. Although delicate and time consuming, the handcraft (often referred to as a scotch tape technique) provides crystals of high structural and electronic quality, which can currently reach sizes of a couple of mm. It is likely to remain the technique of choice for basic research and making proof-of-concept devices in the foreseeable future. Instead of cleaving graphite manually, it is also possible to automate the process by employing, for example, ultrasonic cleavage (*5*). This leads to stable suspensions of submicron graphene crystallites (Fig. 1B), which can then be used to make polycrystalline films and composite materials (*5,6*). Conceptually similar is the ultrasonic cleavage of chemically "loosened" graphite, in which atomic planes are partially detached first by intercalation, making the sonification more efficient (*6*). The sonification allows graphene production on industrial scale.

The alternative route is to start with graphitic layers grown epitaxially on top of other crystals (*7*) (Fig. 1C). This is the 3D growth during which epitaxial layers remain bound to the underlying substrate and the bond-breaking fluctuations suppressed. After the epitaxial structure is cooled down, one can remove the substrate by chemical etching. Technically, this is similar to making, for example, SiN membranes but one-atom-thick crystals were deemed impossible to survive, and no one tried this route until recently (*8-10*). The isolation of epitaxial monolayers and their transfer onto weakly binding substrates (*2*) may now seem obvious but it was realized only last year (*9,10*). With progress continuing apace, the production of graphene wafers looks as a done deal – imagine the following technology. Let us start with a tungsten (011) wafer of many inches in diameter and epitaxially grow a thin Ni (111) film on top (*11*). This is to be followed by chemical vapor deposition of a carbon monolayer (the growth of graphene on Ni can be self-terminating with little lattice mismatch) (*7,11*). In this manner, wafer-scale single crystals of graphene (chemically bound to Ni) have been grown (*11*). A polymer or another film can then be deposited on top, and Ni is etched away as a sacrificial layer leaving a graphene monolayer on an insulating substrate and the expensive W wafer ready for another round. The full cycle has not yet been demonstrated and will probably differ from the gedanken one outlined above (e. g., Cu can be used instead of Ni). Nonetheless, wafers of continuous few-layer graphene have already been grown on polycrystalline Ni films and transferred onto plastic and Si wafers (*9,10*). Importantly, these films exhibit carrier mobility $\mu$ of up to 4,000 cm$^2$/Vs (*10*), close to that of cleaved graphene, even before the substrate material, growth and transfer procedures have been optimized.

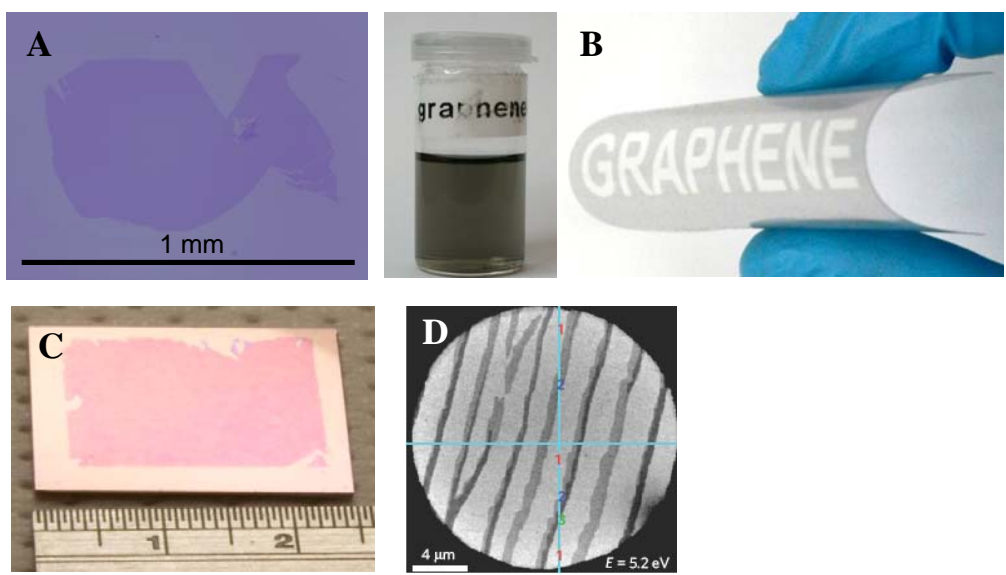

Fig. 1. Making graphene. (**A**) Large graphene crystal prepared on an oxidized Si wafer by the scotch tape technique (courtesy of Graphene Industries Ltd). (**B**) Suspension of microcrystals obtained by ultrasound cleavage of graphite in chloroform (left). Such suspensions can be printed on various substrates. The resulting films are robust and remain highly conductive even if folded (right; courtesy of R. Nair, Manchester). (**C**) First graphene wafers. Polycrystalline one-to-five-layer films grown on Ni and transferred onto a Si wafer (courtesy of A. Reina and J. Kong, MIT). (**D**) State-of-the-art SiC wafer with atomic terraces covered by a graphitic monolayer (indicated by '1'). Double and triple layers ('2' and '3') grow at the steps (*12*).

Where does this leave graphitic layers grown on SiC (*4,12*) (Fig. 1D)? These have been considered as a champion route to graphene wafers for electronics applications, mostly because SiC automatically provides an insulating substrate. First of all, one must distinguish between two principally different types of "graphene on SiC". One is single and double layers grown on the Si-terminated face, and the other is "multilayer epitaxial graphene" that rapidly grows on the carbon face (*4,12*). In the former case, carbon layers are bound to the substrate sufficiently weakly to retain graphene's linear spectrum away (>0.2 eV) from the charge neutrality point (NP) (*13*). However, interaction with the substrate induces strong doping (~$10^{13}$ cm$^{-2}$) and significant spectral disorder at low energies (*13,14*). The crystal quality and coverage homogeneity for the Si-face films have recently improved (*12*), and $\mu$ values approach those for graphene transferred from Ni. As for the carbon face, its epitaxial multilayers should probably be referred to as turbostratic graphene because they are rotationally disordered (no Bernal stacking) and separated by a distance slightly larger than that in graphite (*4,15*). Turbostratic graphene exhibits the Dirac-like spectrum of free-standing graphene, little doping and exceptionally high electronic quality ($\mu$ ~250,000 cm$^2$/Vs at room $T$) (*15*). These features can be attributed to weak electronic coupling between inner layers, their protection from the environment by a few outer layers and the absence of microscopic corrugations (*2,8*). Because an external electric field is screened within just a couple of near-surface layers, turbostratic graphene probably offers limited potential for electronics but is interesting from other perspectives and, especially, for fundamental studies close to NP.

Whichever way one now looks at the prospects for graphene production in bulk and wafer-scale quantities, those challenges that looked so daunting just two years ago, have suddenly shrunk if not evaporated, thanks to the recent advances in growth, transfer and cleavage techniques.

**Quantum Update**
The most explored aspect of graphene physics is its electronic properties. Despite being recently reviewed (*2-4*), this subarea is so important that it necessitates a short update. From the most general perspective, several features make graphene's electronic properties truly unique and different from those of any other known condensed matter system. The first and most discussed is of course graphene's electronic spectrum. Electron waves propagating through the honeycomb lattice completely lose their effective mass, which results in quasiparticles that are described by a Dirac-like equation rather than the Schrödinger equation (*2-4*). The latter – so successful for the understanding of quantum properties of other materials – does not work for graphene's charge carriers with zero rest mass. Figure 2 provides a visual summary of how much our quantum playgrounds expanded due to the experimental discovery of graphene. Second, electron waves in graphene propagate within a layer that is only one atom thick, which makes them accessible and amenable to various scanning probes, as well as sensitive to the proximity of other materials such as high-$\kappa$ dielectrics, superconductors, ferromagnetics, etc. This feature offers many enticing possibilities in comparison with the conventional 2D electronic systems (2DES). Third, graphene exhibits an astonishing electronic quality. Its electrons can cover submicron distance without scattering, even in samples placed on an atomically rough substrate, covered with adsorbates and at room *T*. Fourth, due to the massless carriers and little scattering, quantum effects in graphene are robust and can survive even at room *T*.

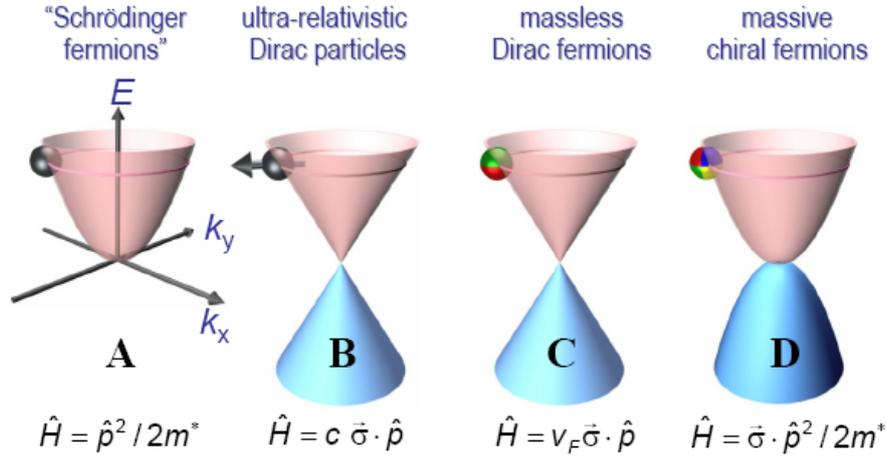

Fig. 2. Quasiparticle zoo. (**A**) Charge carriers in condensed matter physics are normally described by the Schrödinger equation with an effective mass $m^*$ different from the free electron mass ($p$ is the momentum operator). (**B**) Relativistic particles in the limit of zero rest mass follow the Dirac equation, where $c$ is the speed of light and $\sigma$ is the Pauli matrix. (**C**) Charge carriers in graphene are called massless Dirac fermions and described by a 2D analogue of the Dirac equation with the Fermi velocity $v_F \approx 1\times10^6$ m/s playing the role of the speed of light and a 2D pseudospin matrix $\sigma$ describing two sublattices of the honeycomb lattice (*3*). Similar to the real spin that can change its direction between, say, left and right, the pseudospin is an index that indicates on which of the two sublattices a quasiparticle is located. The pseudospin can be indicated by color (say, red and green). (**D**) Bilayer graphene provides us with yet another type of quasiparticles that have no analogies. They are massive Dirac fermions described by a rather bizarre Hamiltonian that combines features of both Dirac and Schrödinger equations. The pseudospin changes its color index 4 times as it moves between four carbon sublattices (*2-4*).

The initial studies of graphene's electronic properties were focused on the analysis of what new physics could be gained by using the Dirac equation within the standard condensed matter formalism (*2-4*). This "recycling" of quantum electrodynamics for the case of graphene has quickly led to the understanding of the half-integer quantum Hall effect (QHE) and the predictions of such phenomena as Klein tunneling, zitterbewegung, the Schwinger production (*16*), supercritical atomic collapse (*3,17*) and Casimir-like interactions between adsorbates on graphene (*18*). As for experiment, only the Klein tunneling has been verified in sufficient detail (*19,20*). Furthermore, transport properties of real graphene devices have turned out to be much more complicated than theoretical quantum electrodynamics, and some basic questions about graphene's electronic properties still remain to be answered. For example, there is no consensus about the scattering mechanism that currently limits $\mu$, little understanding of transport properties near NP [especially, on zero Landau level (*21*)] and no evidence for many predicted interaction effects.

In the near term, much of this research will continue being driven by our knowledge about other low-D systems and using the "recycling" of the known issues and phenomena. Graphene-based quantum dots (*22,23*), p-n junctions (*19,20*), nanoribbons (*23-25*), quantum point contacts (*22*) and, especially, magnetotransport near NP have not received even a fraction of the attention they deserve. Also, it is easy to foresee the revisiting of lateral superlattices, magnetic focusing, electron optics and many interference and ballistic effects studied previously in the conventional 2DES (*26*), which hopefully can either be more spectacular in graphene or clarify its physics. Among other usual suspects is electro- and magneto- optics where graphene offers many unexplored opportunities.

A truly unique feature is that graphene is structurally malleable, and its electronic, optical and phonon properties can be strongly modified by strain and deformation (*27*). For example, strain allows one to create local gauge fields (*3*) and even alter graphene's band structure. Research on bended, folded and scrolled graphene is also gearing up. Furthermore, graphene and turbostratic graphene offer a dream playground for scanning probe microscopy and, personally, I look forward to many possible experiments that can observe supercritical screening, detect local magnetic moments, map wavefunctions in quantizing fields, etc. Further down the line are interaction effects in split bilayers. Experimentally challenging, but it may bring up physics even more spectacular than that in the other 2DES (*28*). Last but not least on my wish list is the fractional QHE. Its possibility has already been tormenting graphene researchers who occasionally observe plateau-like features at fractional fillings, only to find them irreproducible for different devices.

The above sketchy agenda may take many years to complete, and the speed of developments will crucially depend on progress with growing wafers and improving samples' quality. Inch-size wafers with million range $\mu$ can no longer be dismissed as "graphene dreams" and, when this happens, many-body phenomena and new physics that cannot even be envisaged at this stage are likely to emerge.

**Chemistry Matters**

Graphene is an ultimate incarnation of the surface: It has two faces with no bulk left in between. While this surface's physics is currently at the center of attention, its chemistry remains largely unexplored. What we have so far learned about graphene chemistry is that, similar to the surface of graphite, graphene can adsorb and desorb various atoms and molecules (for example, $NO_2$, $NH_3$, K and OH). Weakly attached adsorbates often act as donors or acceptors and lead to changes mostly in the carrier concentration so that graphene remains highly conductive (*29*). Other adsorbates like $H^+$ or $OH^-$ give rise to localized ("mid-gap") states close to NP, which results in poorly conductive derivatives such as graphene oxide (*6*) and "single-sided graphane" (*30*). Despite the new names, these are not new chemical compounds but the same graphene randomly decorated with adsorbates. Thermal annealing or chemical treatment allows the reduction of graphene to its original state with relatively few defects left behind (*30*). This reversible dressing up and down is possible due to the robust atomic scaffold that remains intact during chemical reactions.

Within this surface science perspective, graphene chemistry looks similar to that of graphite, and the latter can be used for guidance. There are principal differences too. First, chemically-induced changes in graphene's properties are much more pronounced because of the absence of an obscuring contribution from the bulk (*29*). Second, unlike graphite's surface, graphene is not flat, but typically exhibits nm-scale corrugations (*8*). The associated strain and curvature can markedly influence local reactivity. Third, reagents can attach to both graphene faces, and this alters the energetics allowing chemical bonds that would be unstable if only one surface were exposed (*31*).

An alternative to the surface chemistry perspective is to consider graphene as a giant flat molecule (as first suggested by Linus Pauling). As any other molecule, graphene can partake in chemical reactions. The important difference between the two viewpoints is that in the latter case adsorbates are implicitly assumed to attach to the carbon scaffold in a stoichiometric manner, that is, periodically rather than randomly. This should result in new 2D crystals with distinct electronic structures and different electrical, optical, chemical, etc. properties. The first known example is graphane, a 2D hydrocarbon with one hydrogen atom attached to every site of the honeycomb lattice (*30,31*). Many other graphene-based crystals should be possible because adsorbates are likely to self-organize into periodic structures, similar to the case of graphite well known for its surface superstructures. Instead of doping with atomic hydrogen as in graphane, $F^-$, $OH^-$ and many functional groups look viable candidates in the search for novel graphene-based 2D crystals.

Graphene chemistry is likely to play an increasingly important role in future developments. For example, stoichiometric derivatives offer a way to control the electronic structure, which is of interest for many applications including electronics. Chemical changes can probably be induced even locally. Imagine then an all-graphene circuitry with interconnects made from pristine graphene whereas other areas are modified to become semiconducting and allow transistors. Disordered graphene-based derivatives should not be overlooked either. They can probably be referred to as functionalized graphene so that it is suitable for specific applications. "Graphene paper" is a spectacular example of how important such functionalization could be (Fig. 3). If it is made starting with a suspension of non-functionalized flakes (*5*), the resulting material is porous and extremely fragile and can float in air like a feather. However, the same paper made of graphene oxide is dense, stiff and strong (*6,32*). In the latter case, the functional groups bind individual sheets together, which results in a microscopic structure not dissimilar to that of nacre known for its strength. Instead of aragonite bound in nacre by biopolymer glue, graphene oxide laminate and, especially, its reduced version (*32*) make use of atomic-scale stitching of the strongest known nanomaterial.

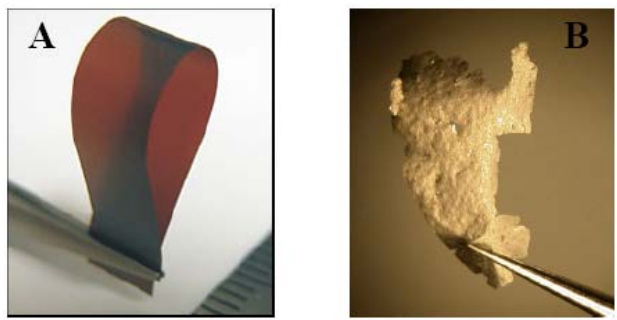

Fig. 3. Graphene derivatives. (**A**) Graphene-oxide laminate is tough, flexible, transparent and insulating (*6*). (**B**) Paper made in the same way as **A** but starting from graphene suspension (*5*) is porous, fragile, opaque and metallic (courtesy of R. Nair, Manchester).

Despite a cornucopia of possible findings and applications, graphene chemistry has so far attracted little interest from professional chemists. One reason is of course that graphene is neither a standard surface nor a standard molecule. However, the main obstacle has probably been the lack of samples suitable for traditional chemistry. The recent progress in making graphene suspensions (*5,6*) has opened up a way to liquid-phase chemistry, and, hopefully, the professional help that graphene researchers have long been waiting for is now coming.

**Sleeping Beauty**
It is customary these days to start reports on graphene by referring to it as a "unique electronic system". In my opinion, this statement belittles what graphene is actually about. 2DES and even Dirac-like quasiparticles were known before but one-atom-thick materials were not. In this respect, graphene has founded a league of its own but little remains known about its other, non-electronic properties. The situation is now rapidly changing and this brings beautiful new dimensions into graphene research.

Last year, the first measurements of graphene's mechanical and thermal properties were reported. It exhibits a breaking strength of ≈40 N/m, reaching the theoretical limit (*33*). Record values for room-$T$ thermal conductivity (≈5,000 W/mK) (*34*) and Young's modulus (≈1.0 TPa) (*33*) were also reported. Graphene can be stretched elastically by as much as 20% as no other crystal can (*33*). These observations were partially expected on the basis of previous studies of carbon nanotubes and graphite, which are structurally made of graphene sheets. Somewhat higher values observed in graphene can be attributed to the virtual absence of crystal defects in samples obtained by micromechanical cleavage. Even more intriguing are those findings that have no analogues. For example, unlike any other material, graphene shrinks with increasing $T$ at all $T$ due to membrane phonons dominating in 2D (*35*). Also, graphene exhibits simultaneously high pliability (folds and pleats are commonly observed) and brittleness (it fractures like glass at high strains (*36*)). The notions are an oxymoron but graphene somehow combines both properties. Equally unprecedented is the observation that the one-atom-thick film is impermeable to gases, including helium (*37*). When wafers become available, I expect an explosion of interest in (bio) molecular and ion transport through graphene and its membranes with designer pores.

Speaking of non-electronic properties, we do not even know such basic things about graphene as how it melts. Neither the melting $T$ nor even the order of the phase transition is known. Ultra-thin films are known to exhibit melting temperatures that rapidly decrease with decreasing thickness. Thermodynamics of 2D crystals in a 3D space could be very different from that of thin films and resemble more the physics of soft membranes. For example, melting can occur through generation of defect pairs and be dependent on the lateral size, similar to the Kosterlitz-Thouless transition. Experimental progress in studying graphene's thermodynamic properties has been hindered by small sizes of available crystals but the situation may change soon. On the other hand, theoretical progress is likely to remain slow because small sizes have proven to be a problem also in molecular dynamics and other numerical approaches, which struggle to grasp the underlying physics when studying crystals of only a few nm in size.

**Grandeur and Plainness**

Potential applications of graphene were discussed in ref. (*2*) and, during the last 2 years, significant progress has been made along many lines penciled in there. The major difference between now and then is the advent of graphene's mass production technologies. This has dramatically changed the whole landscape by making the subject of applications less speculative and allowing the development of new concepts unimaginable earlier.

Most of the buzz is currently around graphene's long-term prospects in computer electronics. Immediate, but often mundane, applications are least discussed and remain unnoticed even within parts of the graphene community. An extreme example of the former is an idea about graphene becoming the base electronic material "beyond the Si age". Although this possibility cannot be ruled out, it is so far beyond the horizon that it cannot be assessed accurately either. At the very least, graphene-based integrated circuits require the conducting channel to be completely closed in the off state. Several schemes have been proposed to deal with graphene's gapless spectrum and, recently, nanoribbon transistors with large on-off current ratios at room $T$ were demonstrated (*22,25*) (Fig. 4A). Nevertheless, the prospect of "graphenium inside" remains as distant as ever. This is not because of graphene shortfalls but rather because experimental tools to define structures with atomic precision are lacking. More efforts in this direction are needed but the progress is expected to be painstakingly slow and depend on technological developments outside the research area.

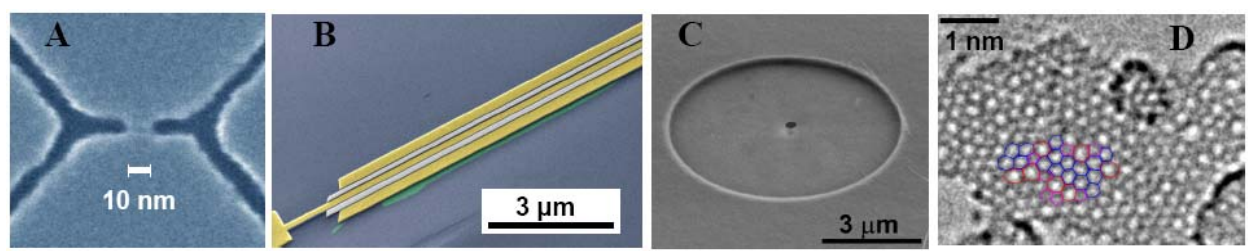

Fig. 4. From dreams to reality. (**A**) Graphene nanoribbons of sub-10-nm scale exhibit the transistor action with large on-off ratios (*22,25*). Scanning electron micrograph shows such a ribbon made by electron-beam lithography (*22*). Control of ribbon's width and its edge structure with atomic precision remains a daunting challenge on the way towards graphene-based electronics. (**B**) All the fundamentals are in place to make graphene-based HEMTs. The false colour micrograph shows the source and drain contacts in yellow, two top gates in light grey and graphene underneath in green (*38*). Courtesy of Y. Lin, IBM. (**C**) Graphene-based NEMS. Shown is a drum resonator made from a 10-nm-thick film of reduced graphene oxide, which covers a recess in a Si wafer (*32*). (**D**) Ready to use: graphene membranes provide an ideal support for TEM. The central part is a monolayer of amorphous carbon. Graphene itself shows in this image only as a grey background (see the top part). Carbon atoms in the amorphous layer appear dark and make a random array of pentagons, hexagons and heptagons as indicated by colour lines (courtesy of J. C. Meyer, A. Chuvilin and U. Kaiser, Ulm). Individual oxygen atoms clearly visible on graphene were also reported (*36*).

An example to the contrary is the use of graphene in transmission electron microscopy (TEM). It is a tiny niche application but it is real. Single-crystal, one atom thick and low atomic mass membranes provide the best imaginable support for atomic resolution TEM. With micron-sized crystallites now available in solution (*5*) for their cheap and easy deposition on standard grids and with films transferrable from metals (*9,10*) onto such grids, graphene membranes are destined to become a routine TEM accessory (Fig. 4D).

The space between graphene dreams and immediate reality is packed with applications. The one that is neither grand nor mundane is individual ultra-high frequency analogue transistors (Fig. 4B). This area is currently dominated by GaAs-based devices known as high electron mobility transistors (HEMT), which are widely used in communication technologies. Graphene offers a possibility to extend HEMT's operational range into THz frequencies. The fundamentals allowing this are well known: Graphene exhibits room-$T$ ballistic transport such that the charge transit between source and drain contacts takes only 0.1ps for a typical channel length of 100 nm. Also, gate electrodes can be placed as close as several nm above graphene, which allows shorter channels and even quicker transit. Although graphene's gapless spectrum leads to low on-off ratios of 10 to 100, they are considered sufficient for the analogue electronics. The progress towards graphene HEMTs is hindered by experimental difficulties in accessing the microwave range. Only recently, the first frequency tests of graphene transistors were reported (*38*). Long channels and low mobility in these experiments limited the cut-off frequencies to less than 30 GHz (*38*), well below the operational range of GaAs-based HEMTs. However, the observed scaling of the operational frequency as a function of the channel length and $\mu$ indicates that the THz range is accessible (*38*). With graphene wafers in sight, these efforts are going to intensify, and HEMTs and other ultrahigh-frequency devices such as switches and rectifiers have a realistic chance to reach the market.

**Sitting on a Graphene Mine**
There has been an explosion of ideas that suggest graphene for virtually every feasible use. This is often led by analogies with carbon nanotubes that continue to serve as a guide in searching for new applications. For example, graphene powder is considered to be excellent filler for composite materials (*6*). Reports have also been made on graphene-based supercapacitors, batteries and field emitters, but it is too early to say whether graphene is able to compete with the other materials, including nanotubes. Less expectedly, graphene has emerged as a viable candidate for the use in optoelectronics (*10,39*). Suspensions offer an inexpensive way for making graphene-based coatings by spinning or spraying (Fig. 1B). An alternative is the transfer of films grown on Ni (*9,10*). These coatings are often suggested as a competitor for indium tin oxide (ITO) that is the industry standard in such products as solar cells, LCD displays, etc. However, graphene films exhibit resistivity of several hundred Ohms for the standard transparency of ~80% (*9,10,39*). This is 2 orders higher than for ITO and unacceptable in many applications (for example, solar cells). It remains to be seen whether the conductivity can be improved to the required extent. Having said that, graphene coatings also offer certain advantages over ITO. They are chemically stable, robust and flexible and can even be folded, which gives them a good chance of beating the competition in touch screens and bendable applications.

There is also fast growing interest in graphene as a base material for nanoelectromechanical systems (NEMS) (*32,40*). Graphene is ultimately light and stiff, which are the essential characteristics sought in NEMS for sensing applications. Graphene-based resonators offer low inertial masses, ultra-high frequencies and, in comparison with nanotubes, low-resistance contacts that are essential for matching the impedance of external circuits. Graphene membranes have so far shown quality factors of ~100 at 100 MHz frequencies (*40*). Even more encouraging are data for drum resonators made from reduced graphene oxide films (*32*). These nm-thick polycrystalline NEMS (Fig. 4C) exhibit high Young's moduli (comparable to those of graphene) and quality factors of ~4,000 at room $T$. Importantly, the films can be produced as wafers and then processed by standard microfabrication techniques. Further developments (increasing the frequency and improving quality factors) should allow graphene NEMS to assail such tantalizing challenges as inertial sensing of individual atoms and the detection of zero-point oscillations.

Among other applications that require mentioning are labs-on-chips (electronic noses) and various resistive memories. The high sensitivity of graphene to its chemical environment is well acknowledged, after sensors capable of detecting individual gas molecules were demonstrated (*29*). The prospect of graphene wafers has put this curiosity-driven record into another perspective. Imagine an array of graphene devices each functionalized differently to be able to react to different chemicals or bio-molecules. Such functionalization has been intensively researched for the case of carbon nanotubes, and graphene adds the possibility of mass-produced arrays of identical devices. Furthermore, there are several enticing reports on non-volatile memories in which graphene-based wires undergo reversible resistance switching by, for example, applying a sequence of current pulses (*41,42*). The underlying mechanism remains largely unknown but such nm-scale switches present an attractive alternative to phase-change memories and deserve further attention. Reports on graphene-ferroelectric memories (*43*) are also encouraging, especially due to the basic simplicity of their operation.

**More Room in the Flatland**
Graphene has rapidly changed its status from being an unexpected and sometimes unwelcome newcomer to a rising star and to a reigning champion. The professional skepticism that initially dominated the attitude of many researchers (including myself) with respect to graphene applications is gradually evaporating under the pressure of recent developments. Still, it is the wealth of new physics – observed, expected and hoped for – which is driving the area for the moment. Research on graphene's electronic properties is now matured but is unlikely to start fading any time soon, especially because of the virtually unexplored opportunity to control quantum transport by strain engineering and various structural modifications. Even after that, graphene will continue to stand out as a truly unique item in the arsenal of condensed matter physics. Research on graphene's non-electronic properties is just gearing up, and this should bring up new phenomena that can hopefully prove equally fascinating and sustain, if not expand, the graphene boom.

44. This work was supported by EPSRC (UK), Office of Naval Research, and Air Force Office of Scientific Research. The author acknowledges many helpful comments of Irina Grigorieva, Antonio Castro Neto, Allan MacDonald, Philip Kim and Kostya Novoselov.